\documentclass[twocolumn]{revtex4}
\usepackage{epsf,graphicx}
\begin{document}
\newcommand{\rme}{{\mathrm{e}}}
%
\newcommand{\fig}[2]{\epsfxsize=#1\epsfbox{#2}}
\newlength{\bilderlength}
\newcommand{\bilderscale}{0.21}
\newcommand{\storebilderscale}{\bilderscale}
\newcommand{\bilderskip}{\hspace*{0.8ex}}
\newcommand{\textdiagram}[1]{%
\renewcommand{\bilderscale}{0.27}%
\diagram{#1}\renewcommand{\bilderscale}{\storebilderscale}}
\newcommand{\diagram}[1]{%
\settowidth{\bilderlength}{\bilderskip%
\includegraphics[scale=\bilderscale]{./pictures/#1.eps}\bilderskip}%
\parbox{\bilderlength}{\bilderskip%
\includegraphics[scale=\bilderscale]{./pictures/#1.eps}\bilderskip}}
\newcommand{\Diagram}[1]{%
\settowidth{\bilderlength}{%
\includegraphics[scale=\bilderscale]{./pictures/#1.eps}}%
\parbox{\bilderlength}{%
\includegraphics[scale=\bilderscale]{./pictures/#1.eps}}}

\author{Pierre Le Doussal{$^1$} and Kay J\"org Wiese{$^2$}}
\affiliation{{$^1$}CNRS-Laboratoire de Physique Th{\'e}orique de
l'Ecole Normale Sup{\'e}rieure,
24 rue Lhomond 75231 Paris, France\\
$^2$ Institute for Theoretical Physics, University of
California at Santa Barbara, Santa Barbara, CA 93106-4030, USA}
\title{\sffamily Functional renormalization group at large $N$ for random manifolds}
\begin{abstract}\smallskip
We introduce a method, based on an exact calculation of the
effective action at large $N$, to bridge the gap
between mean field theory and renormalization in complex systems.
We apply it to a $d$-dimensional manifold in a random potential
for large embedding space dimension $N$. This yields a functional
renormalization group equation valid for any $d$, which contains
both the $O(\epsilon=4-d)$ results of Balents-Fisher and some of the
non-trivial results of the Mezard-Parisi solution thus shedding
light on  both. Corrections are computed at order
$O(1/N)$. Applications to the problems of KPZ, random field and
mode coupling in glasses are mentioned.
\end{abstract}
\maketitle

The random manifold problem, i.e.\ the behavior of an elastic
interface in a random potential is important for many
experimental systems and still offers a considerable theoretical
challenge \cite{book_young}. It is the simplest example of a class
of disordered systems, including random field magnets, where
the so called dimensional reduction \cite{dimred2} renders
conventional perturbation theory trivial and useless. It also
provides powerful analogies, via mode coupling theory, to complex
systems such as structural glasses \cite{book_young}. Two
analytical approaches have been devised so far, in limits where
the problem {\it appears} to simplify while remaining non-trivial:
the functional renormalization group (FRG)
\cite{fisher_functional_rg,balents_fisher,twolooplarkin} and the
(mean field) replica gaussian variational method (GVM)
\cite{mezard_parisi}, together with their dynamical versions
\cite{mfdyn,frgdyn,frgdynT}. The FRG is hoped to be 
controlled for small $\epsilon=4-d$, where $d$ is the 
internal dimension of the interface 
(parameterized by a $N$ component vector $\vec u(x)$ in the embedding space).
It follows the second cumulant of the random potential $R(u)$ under coarse graining,
which becomes non-analytic at $T=0$ beyond the Larkin scale. The GVM
approximates the replica measure by a replica symmetry broken
(RSB) gaussian, equivalently, the Gibbs measure for $u$ as a
random superposition of gaussians \cite{mezard_parisi}, and
is claimed to be exact for $N=\infty$. Computing next order corrections
is fraught with difficulties \cite{cyrano},
and it is still unclear in which sense both methods describe small but {\it finite}
$\epsilon$ or $1/N$. The GVM for instance predicts a transition
for $d=0$ which must disappear for any finite $N$. We found recently
\cite{twoloop,chauve_pld} that higher loop FRG equations for
$R(u)$ at $u \neq 0$ contain non-trivial, potentially ambiguous
``anomalous terms'' involving the non-analytic structure of $R(u)$
at $u = 0$. Although a solution was found to two loops
\cite{twoloop,chauve_pld}, the many loop structure 
remains mysterious. 
%
While both methods  circumvent dimensional
reduction by providing a non-perturbative mechanism, the GVM via replica
symmetry breaking and the FRG via the generation of a cusp-like
non-analyticity in $R''(u)$, they are disconcertingly different in spirit.
Physically, however, both capture the 
metastable states beyond the Larkin scale $R_c$ and should thus be
related \cite{balents_rsb_frg}. A useful, quantitative and more general
method, where this connection appears naturally, is still lacking. 

In this Letter we introduce such a method, encompassing both the FRG and
the GVM. Via an exact calculation of the effective action $\Gamma[u]$ at large $N$, we
obtain the FRG $\beta$-function in any $d$ at large $N$. Its detailed
analysis at dominant order, $N=+\infty$, reveals that the FRG {\it exactly}
reproduces (without invoking spontaneous RSB) the non-trivial result of the 
GVM for small overlap. The connections can be clarified using that 
$\Gamma[u]$ also gives the probability distribution of a given mode $u_q$.
$O(1/N)$ corrections are computed, with the aim of understanding
finite but large $N$. Further results, extensions and discussions
will appear in \cite{us_long}.

We start from the partition sum of an interface ${\cal Z}_V=\int
{\cal D}[u]\, \rme^{- {\cal H}_V[u]/T}$ in a given 
sample, with energy:
\begin{equation}\label{}
{\cal H}_V[u] = \int_q \frac{1}{2} (q^2 + m^2) u_{-q} \cdot u_{q}
+ \int_x V_x(u(x)) \end{equation} where $\int_q \equiv \int
\frac{d^d q}{(2 \pi)^d}$, $\int_x \equiv \int d^d x$. The small
confining mass $m$ provides a scale. To obtain a non-trivial large
$N$ limit one defines the scaled field $v=u/\sqrt{N}$ and chooses
the distribution of the random potential $O(N)$ rotationally
invariant, e.g.\ its second cumulant as:
\begin{equation}
\overline{V_x(u) V_{x'}(u')} = R(u-u') \delta_{x x'} = N
B((v-v')^2) \delta_{x x'} \label{correlator}
\end{equation}
in terms of a function $B(z)$. Higher connected cumulants are scaled
as $\overline{V_{x_1}(u_1)\dots V_{x_p}(u_p)}^{\mathrm{conn}} = N
\delta_{x_1,\dots ,x_p} S^{(p)}(v_1,\dots ,v_p)$. Physical observables can be
obtained for any $N$ from the replicated action at $n=0$ with a
source ${\cal Z}[j]= \int {\cal D} [u] {\cal D}[ \chi] {\cal
D}[\lambda] \rme^{ - N {\cal S}[u,\chi,\lambda,j] }$
\begin{eqnarray}
 {\cal S}[u,\chi,\lambda,j] &=&  \frac{1}{2} \int_q  (q^2 + m^2) v_{-q}^a \cdot v_{q}^a \\
&+&  \int_x [ U(\chi_x)
- \frac{1}{2} i \lambda^{ab}_x (\chi^{ab}_x - v^{a}_{x} \cdot v^{b}_{x})
- j^{a}_{x}\cdot v^{a}_{x} ]\nonumber
\end{eqnarray}
where the replica matrix field $\chi_x \equiv \chi^{ab}_x$ has
been introduced through a Lagrange multiplier. The bare
interaction matrix potential $U(\chi) = \frac{- 1}{2 T^2}
\sum_{ab} B(\tilde{\chi}_{ab}) - \frac{1}{3! T^3} \sum_{abc}
S(\tilde{\chi}_{ab}, \tilde{\chi}_{bc}, \tilde{\chi}_{ca}) +
\dots$ depends only on $\tilde{\chi}_{ab}
=\chi_{aa} + \chi_{bb} - \chi_{ab} - \chi_{ba}$ and has a cumulant
expansion in terms of sums with higher numbers of replicas.

The effective action functional is defined as Legendre transform
\cite{footnote0} $\Gamma[u] + {\cal W}[J] = \int J\cdot u$, with
${\cal W} [J] = 
\ln {\cal Z}[j]$, $J=\sqrt{N} j$. Its full calculation is given in
\cite{us_long}. Since $\Gamma[u]$ defines the renormalized 1PI
vertices, its zero momentum limit defines the {\it renormalized
disorder}. Thus we  only need the result (per unit volume) for a
{\it uniform} configuration of the replica field
$u^a_x=u^a=\sqrt{N} v^a$:
\begin{eqnarray}
\tilde{\Gamma}(v) = \frac{1}{L^d N} \Gamma(u) = \frac{1}{2 T} m^2
v_a^2 + \tilde{U}(v v)
\end{eqnarray}
where $v v$ stands for the matrix $v_a\cdot v_b$. We have computed the
two first coefficients of the renormalized disorder in the $1/N$
expansion $\tilde{U} = \tilde{U}^0 + \frac{1}{N} \tilde{U}^1 +
\dots $. Defining the notation
$\partial_{ab} U(\phi) \equiv
\partial_{\phi_{ab}} U(\phi)$ for any matrix $\phi$ with
components $\phi_{ab}$, we find at dominant order
\begin{eqnarray}
\partial_{ab} \tilde U^{0}(v v) &=& \partial_{ab} U(\chi(v)) \\
\chi(v) &=& v v + T \int_{q} [ (q^2 + m^2) \delta + 2 T \partial
U(\chi(v)) ]^{-1}\qquad \label{saddle}
\end{eqnarray}
i.e.\ a self consistent equation
for $\partial \tilde{U}^0(v v)$, which, as we now show, 
contains both the GVM and the FRG.

For simplicity, we now set all bare cumulants except $B$ to zero.  The
above equations contain a huge amount of
information, since they encode the full distribution (i.e.\ all
cumulants) of the renormalized disorder, and are thus quite
non-trivial to analyze. One limit where they ``simplify'' is when $v$
is set to zero, since they then reproduce the Mezard Parisi (MP)
equations \cite{mezard_parisi} with $\chi(v{=}0)_{ab}=\int_k
G_{ab}(k)$. These exhibit spontaneous RSB (with multiple solutions \cite{footnote2})
and are solved by a
hierarchical Parisi ansatz for $\chi(v{=}0)_{ab} =
\chi(v{=}0)({\sf u})$ where $0 \leq {\sf u} \leq 1$ is the overlap
between replicas $a$ and $b$. In the opposite limit of ``strong''
explicit symmetry breaking field (all $v_{ab}\equiv v_a - v_b \neq
0$) we expect that the renormalized disorder $\tilde{U}(v v)$ is
given by a single saddle point and can be
expanded in replica sums in terms of unambiguous renormalized
cumulants, i.e.\ up to a constant:
\begin{eqnarray}
\tilde{U}(v v)  = \frac{- 1}{2 T^2} \sum_{ab}
\tilde{B}(v_{ab}^2) - \frac{1}{3! T^3} \sum_{abc}
\tilde{S}(v_{ab}^2,v_{bc}^2,v_{ca}^2) + \dots \nonumber
\end{eqnarray}
 This is the limit solved here, which we show is
the one natural in the FRG, and amounts (in the RSB picture) to
forcing the manifold in distant states. Work is in progress to
analyze the rich crossover to RSB contained in
(\ref{saddle}), when some of the $v_{ab}$ are set to zero.

\begin{figure}[t]
\centerline{ \fig{8.5cm}{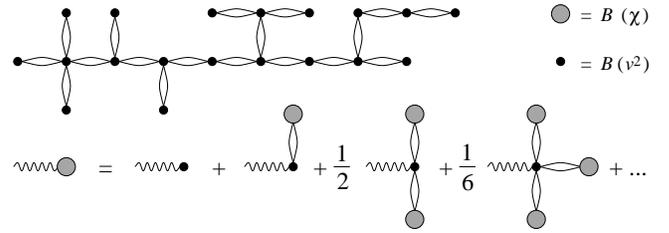} }
\caption{Top: typical $T=0$ contribution to $\tilde B (v_{ab})$. Bottom:
self-consistent equation at leading order for  $\tilde
B' (v_{ab}^{2})=B' (\chi_{ab} )$. The wiggly line denotes
a derivative, and is combinatorially equivalent to choosing one
$B$. At finite $T$ one can attach an additional arbitrary number
of tadpoles to any $B$.} \label{fig1}
\end{figure}

We can now expand, as detailed in \cite{us_long}, any quantity in
(\ref{saddle}) (e.g.\ a replica matrix $M_{ab} = M^0_{ab} + \sum_f
M^1_{abf} + \dots $ and its powers) in sums over an increasing number
of free replica indices. This yields {\it closed} equations for
the second cumulant (with $I_n:=\int_k 1/(k^2 + m^2)^n$ )
\begin{equation}
 \tilde{B}'(v_{ab}^2) = B'(v_{ab}^2 + 2 T I_1 + 4 I_2 (\tilde{B}'(v_{ab}^2) - \tilde{B}'(0)))
 \label{sceq}
\end{equation}
with no other contributions from higher cumulants at any $T$.
This is illustrated
graphically in figure \ref{fig1}. The three replica term
$\tilde{S}$ satisfies a closed equation involving only
$\tilde{B}$, and all  cumulants can be determined iteratively.

The self consistent equation (\ref{sceq}) for the renormalized
disorder can either be inverted directly (done below) or,
equivalently, turned into a FRG equation. We start with the
solution $\tilde{B}(x) = B(x)$ for $m=\infty $ (in presence of a
fixed ultraviolet (UV) cutoff $\Lambda$) and then decrease $m$.
Taking the derivative $m
\partial_m$ and rearranging gives:
\begin{eqnarray}
 m \partial_m \tilde B'(x) = \tilde B''(x) &\!\big[\!&
4 (m \partial_m I_2)(\tilde B'(x)-\tilde B'(0))\nonumber  \\
&& +  2 (m \partial_m T I_1) (1+ 4 I_2 \tilde B''(0))^{-1} \big]\
\ \ \label{rggeneral}
\end{eqnarray}
valid for any $d$ \cite{footnote1}. Since $- \frac{1}{2} m
\partial_m I_1 = m^2 I_2$, (\ref{rggeneral}) has a well defined $\Lambda \to +
\infty$ limit for $d<4$. Then $I_2 = A_d \frac{m^{-
\epsilon}}{\epsilon}$ with $A_d = 2 (4 \pi)^{-d/2} \Gamma[3 -
d/2]$ and we can define the dimensionless function $ b(x) =  4 A_d
m^{4 \zeta - \epsilon} \tilde{B}(x m^{-2\zeta})$, $\zeta$ for now
arbitrary, which satisfies:
\begin{eqnarray}
&&\!\!\! - m \partial_m b(x) = (\epsilon - 4 \zeta) b(x) + 2 \zeta x b'(x) \label{rgscaled}\hfill  \\
&& \qquad\qquad  + {\textstyle \frac{1}{2}} b'(x)^2  - b'(x) b'(0) + T_m b'(x)(
1 + b''(0)/\epsilon)^{-1} \nonumber
\end{eqnarray}
where $T_m= T \frac{4 A_d}{\epsilon} m^{\theta}$, $\theta = d-2 +
2 \zeta$. The FRG equation that we have derived is valid, to
dominant order in $1/N$, {\it in any dimension} $d<4$ and at any
temperature $T$. Restricted to $T=0$ it correctly matches the one
obtained by Balents and Fisher \cite{balents_fisher} at any $N$
but to lowest order in $\epsilon=4-d$. Furthermore, due to the
self consistent equation (\ref{sceq}), it is fully integrable (not
noted in \cite{balents_fisher}). Indeed, (\ref{sceq}) can be
inverted into
\begin{eqnarray}
x &=& m^{2 \zeta} \Phi\Big[ \frac{y}{4 A_d m^{2 \zeta - \epsilon}} \Big] +
\frac1{\epsilon} ( y - y_0 ) - \tilde{T}_m \label{self}
\end{eqnarray}
where $y=-b'(x)$, $y_0 = - b'(0) = - 4 A_d m^{2 \zeta - \epsilon}
B'(0)$, $\Phi$ is the inverse function of $-B' (x)$, i.e.\
$(-B')(\Phi(y))=y$ and 
$\tilde{T}_m = 2 T I_1 m^{2 \zeta}$ ($=T_m/(2-d)$ for $d<2$). That
this is also the general solution of the 
FRG equation can be seen by noting that (\ref{rgscaled}) can be
transformed into a {\it linear} equation for the inverse function
$x(y)$
\begin{equation}
m \partial_m x = (\epsilon - 2 \zeta) y x' + 2 \zeta x - ( y - y_0
) + \frac{ T_m \epsilon x'_0}{1 -  \epsilon x'_0} \label{linear}
\end{equation}
with $x'_0=x'(y_0)$. The general $T=0$ solution of the homogeneous part
reproduces the $\Phi$ term while a particular solution is $x = (y
- y_0)/\epsilon$ using that $m \partial_m y_0 = (2 \zeta -
\epsilon) y_0$.

To analyze the solutions of the large $N$ FRG equation
(\ref{rgscaled}), two approaches are legitimate, corresponding to
different points of view. The first, natural in mean field, is
exact integration. One discovers that
(\ref{sceq}),(\ref{rgscaled}) admit an analytic function as a
solution, given by (\ref{self}), only for $m> m_c$ where $m_c$ is
the Larkin mass (more generally a Larkin scale \cite{footnote1}).
Indeed its second derivative, $\tilde{B}''(0)^{-1} = B''(2 T I_1)^{-1} - 4 I_2$, 
diverges (always for $T=0$, $d<4$, in some cases for $T>0$) when
$m$ is lowered down to $m_c$ (which defines $m_c$). Since
this expression is proportional to the replicon eigenvalue of the
replica symmetric (RS) solution in the GVM, which exists for
$m>m_c$, the generation of a cusp in the FRG exactly coincides at
large $N$ with the instability of the RS solution.

The second approach, natural in the RG \cite{balents_fisher}, is
to view the r.h.s.\ of (\ref{rgscaled}) as the large-$N$ limit of
the true $\beta$-function and to search for a zero. The general
solution for $\theta>0$,  obtained from (\ref{linear}), is
parametrized by $\zeta$:
\begin{eqnarray}
&& x = \frac{y}{\epsilon} - \frac{y_0}{2 \zeta} + \frac{ \epsilon
- 2 \zeta}{ 2 \zeta \epsilon} y_0^{ \frac{\epsilon}{\epsilon - 2
\zeta} } y^{- \frac{2 \zeta}{\epsilon - 2 \zeta} }
\label{fixedpoints}
\end{eqnarray}
for $\zeta>0$ and $\epsilon x = y - y_0 - y_0 \ln(y/y_0)$ for
$\zeta=0$, with $y=-b'(x)$. Here $y_0=-b'(0)$ is a fixed number.
The value of the roughness exponent $\zeta$ \cite{footnote5} is
selected by the decay of $R(u)$ in (\ref{correlator}) at large
$u$, argued to be identical for $B$ and $\tilde{B}$, i.e.\ if
$B'(z) \sim z^{- \gamma}$ one finds $\zeta=\zeta(\gamma) \equiv
(4-d)/2(1+\gamma)$ or $\zeta=0$ for shorter range correlations, to
this order in $1/N$. All the fixed points (\ref{fixedpoints}) have
a cusp $x'(y_0)=0$ (for $\zeta \neq \epsilon/2$) and are expected
to be the physically correct solutions at small $m$.

\begin{figure}[t]
\centerline{ \fig{6 cm}{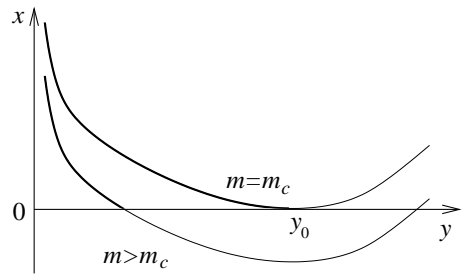} } \caption{The
function $x (y)$ given by (\ref{solu}).}
\label{fig2}
\end{figure}

To show how to reconcile these two results, we study specific
models, the long range (LR) correlations \cite{mezard_parisi}
$B(z) =  \frac{\tilde{g}}{4(\gamma - 1)A_d}  (a^2 + z)^{1 -
\gamma}$ and the short range (SR) gaussian
correlator $B(z)= \frac{\tilde{g}}{4 A_d} e^{- z}$. 
Choosing $\zeta=\zeta(\gamma)$ (\ref{self}) yields:
\begin{eqnarray}
&&  x  = (y/\tilde{g})^{- 1/\gamma} + {\epsilon}^{{-1}} (y - y_0)
- m^{2 \zeta} a^2  - \tilde{T}_m \label{solu}
\end{eqnarray}
(and $\ln(\tilde{g} m^{-\epsilon}/y)$ in the r.h.s.\ for SR
gaussian with $\zeta=0$, $a=0$). As one sees from Fig.\ %
\ref{fig2}, the r.h.s.\ of (\ref{solu}) has a minimum and
decreasing $m$ the curve $x(y)$ cuts the axis $x=0$ closer to the
minimum. It reaches it at $m=m_c$ where the solution acquires a
cusp $b'(0) - b'(x) \approx \sqrt{-2 (\epsilon - 2 \zeta) b'(0)
x}$ and  $m_c^{2 \zeta} a^2 + \tilde{T}_{m_c} = (\tilde{g}
\gamma/\epsilon)^{1/(1+\gamma)} \equiv \tilde{T}_c$, $b'(0) = -
\tilde{g}^{1/(1+\gamma)} (\epsilon/\gamma)^{\gamma/(1+\gamma)}$
($=-\epsilon$ for SR with $m_c^\epsilon=\tilde{g}/\epsilon$).
Although it is a priori unclear how to follow this solution for $m
< m_c$, the following remarkable property indicates how one may
proceed.
If we compute the $\beta$-function, i.e.\ the r.h.s.\ of
(\ref{rgscaled}), using (\ref{solu}) at $m=m_c$ and
$\zeta=\zeta(\gamma)$ we find that it {\it exactly vanishes} for all
$x>0$. Thus, for the potentials studied here, $b(x)$ evolves
according to (\ref{solu}) until $m_c$ where it reaches its fixed
point, and does not evolve for $m<m_c$. This provides
unambiguously a solution beyond the Larkin scale, reconciles the
two approaches and justifies the value
obtained for $\zeta$. The quantity $\tilde{B}''(v^2)^{-1} =
B''(\chi(v))^{-1} - 4 I_2$ plays the role of a replicon 
eigenvalue and remains frozen and positive for $v >0$.

Here, for $N=\infty$, temperature plays only a minor role in the case
where disorder is relevant, i.e.\ for $\theta(\gamma)>0$ (i.e.\ $2 < d < 4$; $d<2$ for
$\gamma < \gamma_c = 2/(2-d)$), which is described by the $T=0$
fixed point (\ref{fixedpoints}). Contrarily to the one loop FRG,
where $b (x)$ remains analytic in a boundary layer $x \sim T_m$ at $T>0$,
here the denominator in (\ref{rgscaled}) (resumming all
orders in $\epsilon$) blows up and a cusp arises. Only
in the marginal case ($\gamma=\gamma_c$; $d=2$ for SR disorder)
 we find a line of fixed points of $(\ref{rgscaled})$  with
$\zeta=(2-d)/2$. For $T>T_{c}$ the disorder is  analytic, given by
(\ref{solu}) for all $m$ ($\tilde{T}_m/\tilde{T}_c \equiv T/T_c$ does
not flow, and $T_c=2 \pi$ in $d=2$ for SR disorder). Below $T_c$ one
recovers a cusp and the $T=0$ fixed point. Finally, for $\gamma >
\gamma_c$ ($d < 2$) no cusp is generated as $m \to 0$ and disorder is
irrelevant, as in the corresponding RS solution of MP. 

Since the FRG aims at obtaining the universal behaviour
at small $m,k$ of the correlation function of
the manifold $\left<v_k^a v_{-k}^b\right> = m^{d-2 \zeta} g(k/m)$,
we can now compute its zero momentum limit $g(0)$, which is
universal in the LR case, and compare with the GVM result $g_{\mathrm{RSB}}(0)$.
Extending \cite{mezard_parisi} in presence of a mass,
one shows that $m\partial_m \sigma_m({\sf u})=0$ and thus the MP self energy is
$\sigma_m({\sf u})=C {\sf u}^{2/\theta-1}$ if ${\sf u}>{\sf u}_m$,
$\sigma_m({\sf u})=\sigma_m(0)$ if ${\sf u}<{\sf u}_m$.
Within the FRG, $g_{\mathrm{FRG}}(0) = - b'(0^+)/(2 A_d)$
and substituting, we discover that 
$g_{\mathrm{FRG}}(0) = T \sigma_m(0)/m^4 < g_{\mathrm{RSB}}(0)$, 
i.e.\ the FRG gives only but exactly the
contribution from the overlaps ${\sf u} < {\sf u}_m$ (the
distant states). This non-trivial information which
 within the MP approach {\it requires} a
{\em full} RSB calculation, is obtained
here {\em without any} RSB. 
Moreover using $m\partial_m \sigma_m({\sf u})=0$ one shows that contributions 
of larger overlaps in the MP result can be obtained by integrating the
$m$ dependent FRG result as
$g_{\mathrm{RSB}}(0) = T[ \sigma_m(0)/m^4 +  \int_{m}^{m_c} \mathrm{d}
\sigma_{m'}(0)/{m'}^4  + m_c^{-2} - m^{-2}] $.

These results can be understood as follows. The effective action 
also describes the probability distribution $P_V(w)$ in a given
environment $V$, of the center of mass of the interface $ w =
\frac{1}{N^{1/2} L^d} \int_x u_x$, i.e.\ one has
$\tilde{\Gamma}[\{w_a \}] = - \lim_{L \to \infty} \frac{1}{N L^d}
\ln \overline{P_V(w_1)\dots  P_V(w_n)}$. Extension of the FRG
beyond the Larkin scale requires giving a meaning to the $u=0^+$ limit.
We find here that what the FRG actually computes (from $b'(0^+)$)
is a second moment of $w$ in presence of a small extra field $\sqrt{N} j_a$
such that all $v_{ab} \neq 0$, i.e.\ an average such that when there are several
states the different replicas are chosen in maximally separated
states (${\sf u}=0$).

In a previous study aiming to connect the RSB solution 
to the FRG \cite{balents_rsb_frg} 
$\overline{\ln P_V(w) \ln P_V(-w)}$ was computed and used to
define a renormalized second cumulant of the disorder. This quantity
is however {\it different} from the one in the FRG, obtained here, and
does not reproduce the second moment of $w$, neither
$g(0)_{\mathrm{RSB}}$, nor $g(0)_{\mathrm{FRG}}$. In addition, since
\cite{balents_rsb_frg} used the unperturbed MP saddle point,
the two calculations focus on different regimes
($v_{ab}^2 \sim 1$ here, $v_{ab}^2 \sim 1/N$ there, $v_a \sim 1$ in both) \cite{leon}.
Work is in progress to connect these regimes, and obtain a more complete
version of the FRG, using our equations (\ref{saddle}) and summing 
over RSB saddle points \cite{footnote2}.

\begin{figure}[t]
\begin{eqnarray*}
\delta B^{(1)}&=&
\!\!\diagram{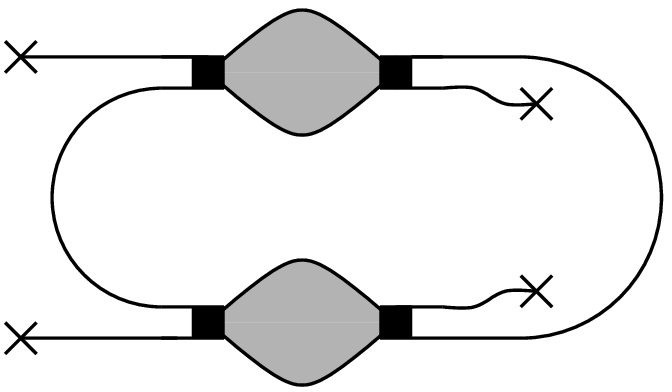}\!\!+\!\!\!\diagram{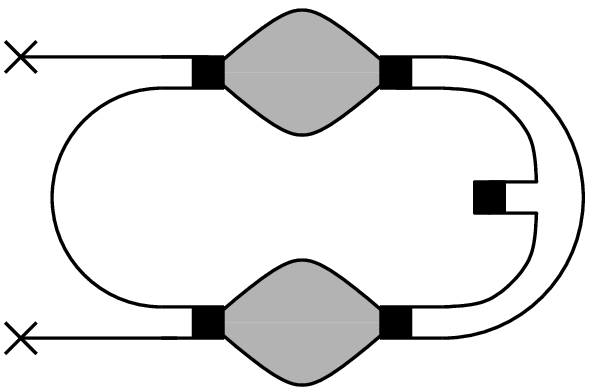}\!\!+\!\!\diagram{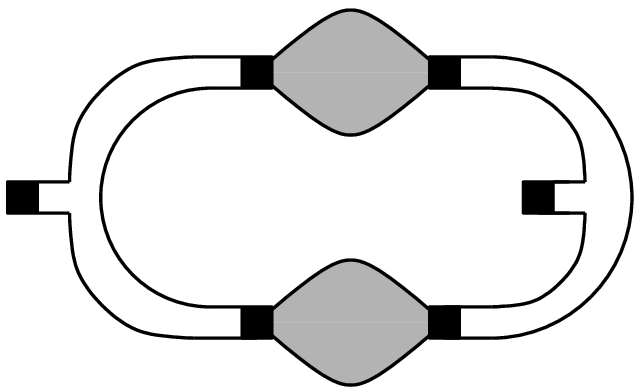}\!\!+\!\!\!\diagram{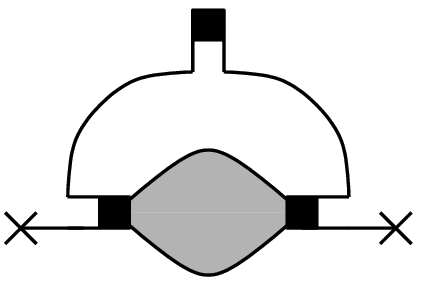}\!\!+\!\!\diagram{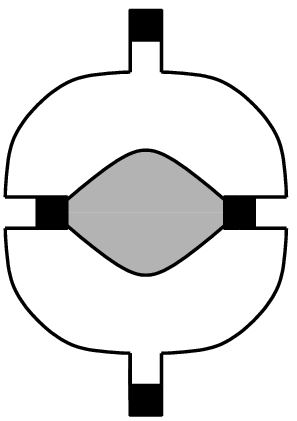}\!\!
\\
&& +T\Big( \!\!\diagram{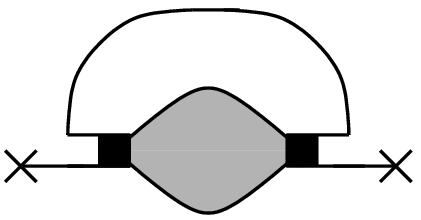} \!\!+ \!\!\diagram{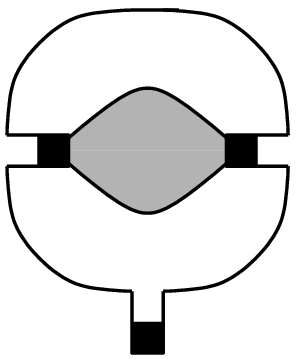} \!\!+
\!\!\diagram{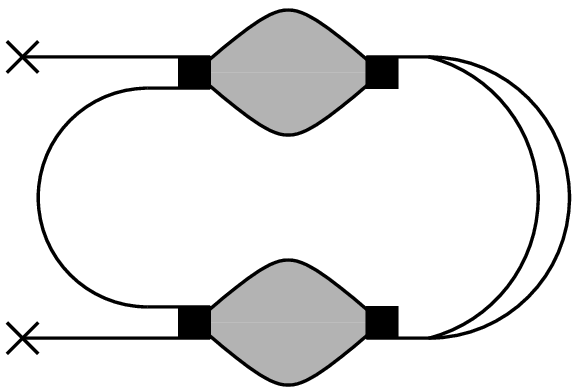} \!\!+ \!\!\diagram{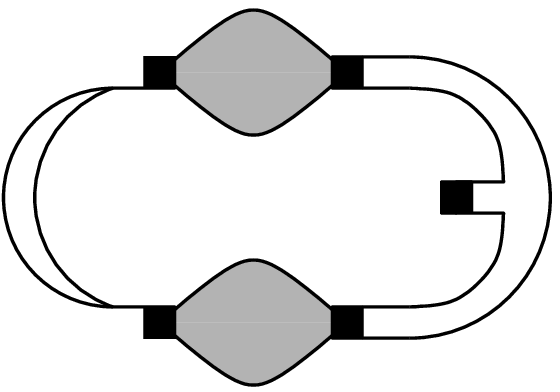}\!\!  \Big)\\
&& + T^{2}\Big( \!\!\diagram{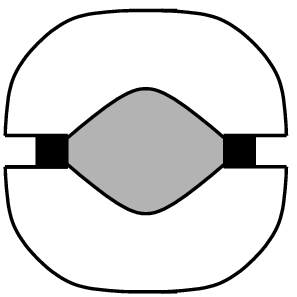} \!\!+ \!\!\diagram{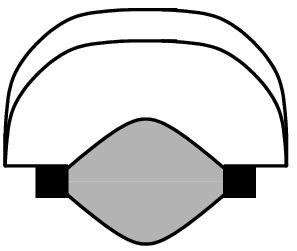}
\!\!+
\!\!\diagram{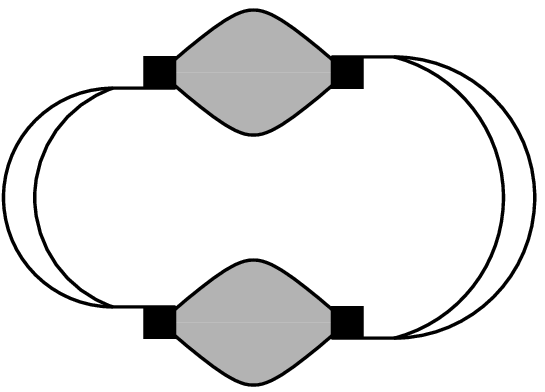}\Big)\\
\diagram{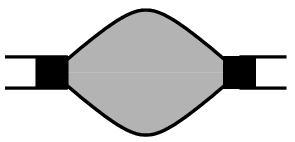}&=&B'' (\chi _{ab})\left(1-4A_{d} I_{2} (p)B'' (\chi
_{ab}) \right)^{-1}\ ,\quad  \diagram{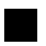}=B(\chi_{ab})
\end{eqnarray*}
\caption{Contribution to the second cumulant at order $1/N$.}
\label{secumul1oN}
\end{figure}%
The FRG approach 
should allow a quantitative study of finite
$N$ beyond possible artifacts of $N=\infty$, and extension of
\cite{balents_fisher} to any $d$. The calculation of 
the correction at order $1/N$ to $\tilde B$ is involved, and
was performed using two complementary methods, a graphical
one, see Fig.\ \ref{secumul1oN}, and the algebraic expansion in number of
replica sums. The resulting expression for the $\beta$-function at
order $1/N$ and $T=0$ is UV-convergent and reads for  $\Lambda \to \infty
$\vspace*{-1cm}
\begin{widetext}\vspace*{-0.5cm}
\begin{eqnarray}
\beta ({b}) &=& \epsilon  {b}   {+} \frac12  {{b}'
} ^{2}{-}
 {b}'    {b}' _0 {+} \frac{1}{NA_{d}}
\int_{p} \Big[
\displaystyle{-} {4  I_0^p I_3^p
    { ( {b}'_0 {-} {b}'   ) }^2 { {b}'' }^2} H_{p}^{{-}2}
{-}{2\epsilon  x {I_0^p}^2
     ( {b}'_0 {-} {b}'   )   {b}''  (1{-}I_{2}b'')
    }H_{p}^{{-}2}
\nonumber \\
&&\displaystyle{+}{\epsilon  I_4^p { ( {b}'_0 {-} {b}'
           ) }^2  {b}''
       ( 2    {+}  ( 2I_{2} {-}  I_2^p  )   {b}''
         ) }H_{p}^{{-}2}
\displaystyle{-}{8\epsilon  x  I_0^p I_3^p
     ( {b}'_0 {-} {b}'   )  { {b}'' }^2
     (   1 {+}I_{2} {b}''   ) }H_{p}^{{-}3}
\displaystyle{+} {2 \epsilon  {I_3^p}^2 { (  {b}'_0 {-}  {b}'   )
        }^2 { {b}'' }^2
     ( 3    {+}  ( 3I_{2} {-}    I_2^p  )   {b}''   )
    }H_{p}^{{-}3}
\nonumber \\
&& \displaystyle{-} {2    {I_0^p}^2 { ( {b}'_0 {-} {b}'   ) }^
     2  {b}'' }H_{p}^{{-}1}
{+}{2\epsilon  x^2  {I_0^p}^2 { {b}'' }^2
     ( 1   {+}  ( I_{2} {+}    I_2^p  )   {b}''   ) }H_{p}^{{-}3}
\Big]\  ,\qquad \displaystyle H_{p}= 1{+} ( I_{2}{-}I_{2}^{p})b''
,\qquad \displaystyle I_{0} ^p = ( 1{+}p^{2})^{-1} \label{betaN}
\end{eqnarray}
\end{widetext}\vspace*{-0.7cm}
with $I_{2}=\epsilon ^{-1}$, $I_2^p=J_{11}^p$, $I_3^p=J_{12}^p$,
$I_4^p=J_{22}^p$ and 
$J_{nm}^{p} = \frac{1}{A_{d}} \int_k (I_0^k)^n  (I_0^{k+p})^m$,
$b'\equiv b'(x)$, $b_0'\equiv b'(0)$, etc...
The corresponding expressions at $T>0$ have been obtained.
Analysis of these formidable expressions is in progress. In
particular, we have not included in (\ref{betaN})  anomalous terms
arising from the non-analytic structure. We have checked that
(\ref{betaN}) non-trivially reproduces the two loop FRG equation
for the $N$-component model of
\cite{twolooplarkin,us_long}.


In summary, via an exact calculation of the effective action 
at large $N$ we have derived  equations valid in any $d$ 
containing both the GVM and the FRG. The FRG and 
its continuation to $m<m_c$ are consistent with 
the main results of the full  and the marginal one step RSB-%
solutions of MP. Since it reproduces the non-trivial small overlap results 
it provides another way to
attack finite $N$. $1/N$ corrections have been obtained.
Our study hints at further connections
between: $1/N$ and thermal boundary layers, RSB, and how to fix the
ambiguities in the anomalous terms in the $\beta$-function. Their
understanding should allow quantitative progress
in the SR case (e.g.\ for $d=1$, equivalent to
 the $N$-dimensional KPZ equation). 
Applications of the method to other complex systems
is in progress. We have also computed the effective action for the
random field $O(N)$-model at large $N$ \cite{us_long}. 
Finally, applications to the 
dynamics offer the hope to go systematically beyond
mode-coupling approximations.


\begin{thebibliography}{999}
\bibitem{footnote0}
The usual definition is the convex envelope $\Gamma[u]=\min_J (J u - W[J])$.
\bibitem{footnote2}
in case of multiple solutions all saddle points should be
summed over $\Gamma[u] = - \ln \sum_\pi e^{- \Gamma_\pi[u]}$.
\bibitem{footnote1}
One can also derive a FRG equation by varying the UV
cutoff from $0$ to $\Lambda$ and a
Larkin scale $\Lambda_c=R_c^{-1}$.
\end{thebibliography}
\end{document}